\documentclass[12pt]{iopart}
\usepackage{amsfonts}
\usepackage{amssymb}
\usepackage[dvips]{graphicx}
\usepackage{colordvi}
\usepackage{times}
\usepackage{subfigure}
\newcommand{\bc}{\begin{center}}
\newcommand{\ec}{\end{center}}
\newcommand{\be}{\begin{equation}}
\newcommand{\ee}{\end{equation}}
\newcommand{\bna}{\begin{eqnarray}}
\newcommand{\ena}{\end{eqnarray}}

\newcommand{\mpaa}{\begin{minipage}[t]{6cm}}
\newcommand{\mpea}{\end{minipage}}
\newcommand{\mpab}{\begin{minipage}[t]{8cm}}
\newcommand{\mpeb}{\end{minipage}}
\newcommand{\mpac}{\begin{minipage}[t]{13cm}}
\newcommand{\mpec}{\end{minipage}}
\newcommand{\mpad}{\begin{minipage}[t]{13cm}}
\newcommand{\mped}{\end{minipage}}
\newcommand{\mpae}{\begin{minipage}[t]{13cm}}
\newcommand{\mpee}{\end{minipage}}
\newcommand{\mpaf}{\begin{minipage}[t]{6cm}}
\newcommand{\mpef}{\end{minipage}}

\newcommand {\bdm} {\begin{displaymath}}
\newcommand {\edm} {\end{displaymath}}

\usepackage{color}
\definecolor{darkblue}{rgb}{0,0,0.6}
\definecolor{darkred}{rgb}{0.7,0,0}
\definecolor{darkgreen}{rgb}{0,0.7,0}
\usepackage{amsfonts}

\newcommand{\bea}{\begin{eqnarray}}
\newcommand{\eea}{\end{eqnarray}}

\def\XXint#1#2#3{{\setbox0=\hbox{$#1{#2#3}{\int}$}
     \vcenter{\hbox{$#2#3$}}\kern-.5\wd0}}

\begin{document}

\title{Fluctuations in quantum one-dimensional thermostatted systems with off-diagonal disorder}

\author{Matteo Colangeli$^1$, Lamberto Rondoni$^{1,2}$}
\address{$^1$Dipartimento di Matematica, Politecnico di Torino, Corso Duca degli Abruzzi
24, I-10129 Torino, Italy\\
$^2$ INFN, Sezione di Torino, Via P. Giura 1, I-10125, Torino, Italy}

\ead{colangeli@calvino.polito.it, lamberto.rondoni@polito.it}

\begin{abstract}
We analyze a one dimensional quantum model with off-diagonal disorder, consisting of a sequence of potential energy barriers whose width is a random variable either uniformly or normally distributed. We investigate how the disorder and the energy distribution (due to a thermostat at room temperature) affect the resulting value of the transmission coefficient, and discuss the structure of the fluctuations of such coefficient at different length scales and the onset of different size-dependent regimes. Our analysis suggests an efficient way to detect tunneling resonances in a regime of off-diagonal disorder.
\end{abstract}

\maketitle

\section{Introduction}
\label{sec:sec0}

Nonequilibrium thermodynamics \cite{dgm} is based on the vast separation between the space and time scales
of the microscopic, mesoscopic and macroscopic physical realms. This separation, which characterizes the condition 
of \textit{local thermodynamic equilibrium} \cite{dgm,Liboff,GibRon}, cannot be realized in small systems,
such as those concerning the modern bio- and nano-technologies. Hence, the standard statistical mechanical 
approach, based on the thermodynamic and hydrodynamic limits, should be extended to treat systems made up
of relatively small numbers of microscopic components. In particular, one should account for the fluctuations 
of the physical properties of such systems, as they can be of the size of the average signals.
In this work we face these issues considering a variant of the Anderson model \cite{Ander}, which is the 
prototype of a disordered solid \cite{Vulp}. In particular, we investigate the role of the microscopic 
disorder on the transmission coefficient of systems with a number $N$ of potential barriers and
in the presence of a thermal reservoir at a given temperature $T$. For this purpose, we consider one 
dimensional systems consisting of a sequence of barriers and wells of randomly chosen widths, under the 
constraint that the sum of the $N$ barrier widths and the total length of the system are fixed and do
not change when $N$ grows. 
Consequently, our model enjoys a purely off-diagonal disorder \cite{TC,SE} which affects the tunneling 
couplings among the wells, but not the energies of the bound states within the wells. This is not the 
case of the original tight-binding model, whose random fluctuations only concern the energy of a bound 
state \cite{Izrailev,CelKap}, introduced by Anderson in his pioneering paper \cite{Ander} on localization 
effects in disordered solids. Furthermore, while in Anderson's model increasing the number of barriers 
corresponds to taking the large system limit, in our case it corresponds to distribute more finely the 
same amount of insulating material within the fixed length of the system. Therefore, our results differ
substantially from those regarding the standard Anderson's model, and can be summarized as follows.
\begin{itemize}
\item There are no localization effects for the equilibrium distribution of energies at room temperature:
positive currents persist even in the large $N$ limit; Furstenberg type theorems do not apply, \cite{Vulp}.
Mathematically, the reason is that the product of the first $N$ random matrices yielding the transmission
coefficient for a given choice of $N$ barriers, changes, in order to preserve the length of the system and 
the sum of the barrier widths, when the $N+1$-th matrix of the $N+1$-th barrier is introduced.  
Physically, our models realize a kind of off-diagonal disorder which leads to delocalization.
\item The value of the transmission coefficient, averaged over the ensemble of disordered configurations, 
is close, for large $N$, to the value corresponding to the ordered sequence of equally spaced barriers 
and wells, which is bounded away from zero.
\item The transmission coefficient increases non-linearly with the temperature parametrizing the
distribution of the energies of the waves entering the system from the charge reservoir, which is also 
a heat bath.
\item There is a scale for $N$, above which the (always positive) transmission coefficient no longer depends 
on the specific realization of the disorder, but still depends on $N$, and there is
another scale above which it no longer depends on $N$. We call ``mesoscopic'' the first, since it corresponds 
to nanoscopic samples, and we call ``macroscopic'' the latter scale, since it corresponds to macroscopic 
nanostructured materials. This means that all realizations of the disorder are equivalent in the $N\to\infty$ limit.
\item At room temperature, the probability distribution function (PDF) of the fluctuations of the transmission
coefficient satisfies a principle of large deviations. Furthermore, the peak of this PDF 
corresponds to the transmission coefficient of the regular realizations.
\item Tunneling resonances for off-diagonal disordered systems in contact with an external bath can
be predicted by investigating the corresponding ordered systems, which are amenable to an analytical 
treatment.
\end{itemize}

\section{The model}
\label{sec:sec1}

\begin{figure}
   \begin{center}
   \includegraphics[width=0.8\textwidth]{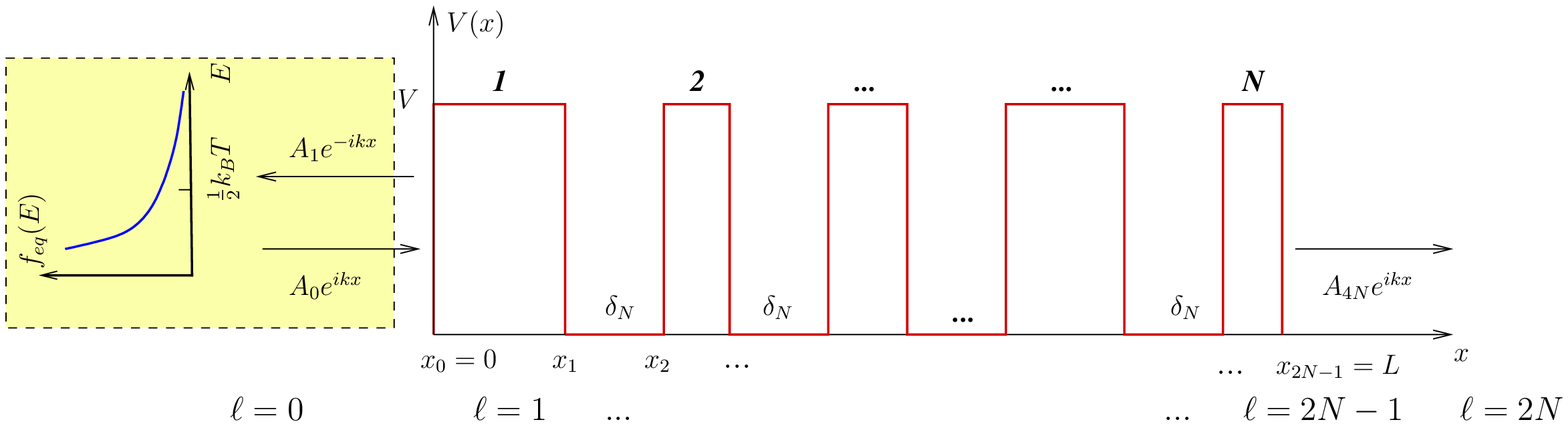}
   \caption{$1$-D Multiple-wells system, consisting of a sequence of $2N+1$ regions: $N$ potential barriers, whose width is uniformly randomly distributed, $(N-1)$ conducting regions of constant width $\delta_N$, and the two boundary regions, on the left and on the right. The left boundary is modeled as a classical thermostat at temperature $T$. Thus, the mean energy of the plane waves entering from the left boundary is given by $\frac{1}{2}k_B T$.}\label{barriers}
   \end{center}
\end{figure}

We consider a model of a macroscopic semiconductor device (the bulk) consisting of an array of $N$ potential barriers and $N-1$ conducting regions (wells), in equilibrium with one electrode acting as an external thermostat at temperature $T=300 K$ (cf. Fig. \ref{barriers}). The barriers have a constant height $V(x)=V$ while their width is either uniformly or gaussian randomly distributed. For any $N$, the widths of the conducting regions take a constant value $\delta_N$.
We denote by $L$ the fixed total length of the sample, by $L^{is}$ the sum of the widths of all the barriers (i.e. the total length of the insulating region), and we also assume that the ratio $\beta$ between insulating and conducting lengths takes the same value for all different sample realizations, so that 
\be 
L=(1+\beta)(N-1)\delta_N  \label{L}
\ee
holds.
This $1$-D microscopic model obeys the steady state Schr\"{o}dinger Equation:

\be
\frac{d^2}{dx^2}\psi=\frac{2m}{\hbar^2}(V-E)\psi, \quad x\in[0,L] \label{se}
\ee
where $m$ is the mass of an electron. The boundary conditions prescribe $A_0>0$ for the amplitude of the plane wave entering from the left boundary and $A_{4N+1}=0$ (no wave enters or is reflected from the right boundary).
The barriers are delimited by a set of $2N$ points, denoted by $x_0=0,...,x_{2N-1}=L$ in Fig. \ref{barriers}, hereafter called \textit{nodes} of discontinuity of the potential. The left boundary consists of a classical thermostat at temperature $T$, from which particles emerge at node in $x_0$ as plane waves, with energies distributed according to the Maxwell-Boltzmann distribution. Differently, no particles come from the vacuum on the right. 

Thus, denoting by $\mathcal{U}_\ell$ the $\ell$-th region, for $\ell\in\{0,2,...,2N\}$, the solutions of eq. (\ref{se}) take the form:

\be
\psi_\ell(x)=\left\{\begin{array}{cc}

A_{2\ell} e^{ikx}+A_{2\ell+1} e^{-ikx} &  \hbox{if $x\in \mathcal{U}_\ell$ and $\ell$ is even (i.e. for $V(x)=0$)} \\ \\
A_{2\ell} e^{-zx}+A_{2\ell+1} e^{zx} &  \hbox{if $x\in \mathcal{U}_\ell$ and $\ell$ is odd (i.e. for $V(x)=V$)} 
               \end{array}\right. \label{psi}
\ee
with $k=\sqrt{2mE}/\hbar$ and $z=\sqrt{2m(V-E)}/\hbar$.

Following \cite{schwabl}, in each of the conducting regions, one may define the steady state \textit{currents} as:

\bea
j_\ell(x)&=&\frac{\hbar}{2mi}\left[\psi_\ell(x)^*\left(\frac{d}{dx}\psi_\ell(x)\right)-(\frac{d}{dx}\psi_\ell(x)^*)\psi_\ell(x)\right]=\nonumber\\
&=&j_\ell^{tr}(A_{2\ell})-j_\ell^{rif}(A_{2\ell+1}) \quad , \label{curr}
\eea
where the $^*$ denotes complex conjugation, $j_\ell^{tr}(A_{2\ell})=\hbar k/m|A_{2\ell}|^2$ denotes the current transmitted from the $(\ell-1)$-th barrier on the left (or, for $\ell=0$, from the thermostat located at the left boundary) and $j_\ell^{tr}(A_{2\ell+1})=\hbar k/m|A_{2\ell+1}|^2$ denotes the current reflected from the $(\ell+1)$-th barrier, cf. Fig. \ref{barriers}. Then, the application of the BenDaniel-Duke boundary conditions \cite{harris}, which require the continuity of $\psi_\ell(x)$ and $\frac{d}{dx}\psi_\ell(x)$ at the nodes, results in the constancy of the value $j_\ell(x)$ across the wells and entails $j_\ell(x)=j_{2N}(x)$, for every even $\ell$. Equation (\ref{curr}), together with eq. (\ref{psi}), leads to the following definition of the transmission coefficient $S$ across the system:
\be
S(N)=\frac{j_{2N}^{tr}(A_{4N})}{j_0^{tr}(A_{0})}=\frac{|A_{4N} |^2}{|A_{0}|^2} \label{J}
\ee
which depends on the parameters of the model, such as the number of barriers $N$, the height of the barriers (which contributes to the parameter $z$) and the particular realization of sequence of barriers. 
In order to numerically compute the coefficient $S$ as a function of the various parameters of the model, it proves convenient to rescale eq. (\ref{se}) with respect to characteristic quantities, in order to rewrite it in a dimensionless form. For this purpose, let us introduce $x=L \tilde{x}$, with $L$ given by (\ref{L}), $E=\tilde{E}E_T$, $\psi(x)=\tilde{\psi}(\tilde{x})(\sqrt{L})^{-1}$, $V=\tilde{V}E_T$, with $E_T=k_B T$ (i.e. twice the mean kinetic energy of the plane waves entering the bulk from the left side). Moreover, by introducing the scalar parameter $\alpha=\hbar^2/(2 m L^2 E_T)$, one obtains the following expression for the dimensionless wavevectors: $\tilde{k}=\sqrt{\alpha}\sqrt{\tilde{E}}$ and $\tilde{z}=\sqrt{\alpha}\sqrt{\tilde{V}-\tilde{E}}$. In the sequel we will refer to the dimensionless quantities and, to this aim, we may omit the tilde symbols, for sake of simplicity.
The dimensionless version of eq. (\ref{se}), then, attains the form:
\be
\frac{d^2}{dx^2}\psi(x)=\alpha^{-1}(V-E)\psi(x), \quad x\in[0,1] \label{se2}
\ee
which is the Schr\"{o}dinger Equation (SE) which will be solved numerically with the aforementioned conditions at the nodes.

\section{The transfer matrix technique}
\label{sec:sec2}

Let us describe our method of solution of the SE (\ref{se2}), which follows Refs. \cite{harris,raz} and is referred to as the Transfer Matrix (TM) technique.
Using eqs. (\ref{psi}), the boundary conditions on the generic $\ell$-th node, with $\ell\in\{0,1,...,2N-1\}$ read as:

\be
\left\{\begin{array}{c}
\psi_\ell(x_\ell)=\psi_{\ell+1}(x_\ell) \\
\psi_\ell'(x_\ell)=\psi_{\ell+1}'(x_\ell)
\end{array} \right. \quad , \label{BD}
\ee
where $x_\ell(x)=\sum_{i=1}^{\ell/2}\lambda_i+\delta_N \ell/2$ if $\ell$ is even, and $x_\ell(x)=\sum_{i=1}^{(\ell+1)/2}\lambda_i+\delta_N (\ell-1)/2$ if $\ell$ is odd, and where $\lambda_i$ denotes the (dimensionless) random width of the $i$-th barrier.
In matrix form, eqs. (\ref{BD}) can be written as:

\bea
\mathbf{M}_0(x_0) \cdot \left( \begin{array}{c}
A_0 \\
A_1 
\end{array} \right)&=&\mathbf{M}_1(x_0)\cdot\left( \begin{array}{c}
A_2 \\
A_3 
\end{array} \right) \nonumber\\
\mathbf{M}_2(x_1) \cdot\left( \begin{array}{c}
A_2 \\
A_3 
\end{array} \right)&=&\mathbf{M}_3\cdot(x_1)\left( \begin{array}{c}
A_4 \\
A_5 
\end{array} \right) \nonumber\\
\mathbf{M}_4(x_2) \cdot\left( \begin{array}{c}
A_4 \\
A_5 
\end{array} \right)&=&\mathbf{M}_5(x_2)\cdot\left( \begin{array}{c}
A_6 \\
A_7 
\end{array} \right) \label{trmat}\\
&\vdots&\nonumber\\
\mathbf{M}_{4N-2}(x_{2N-1}) \cdot\left( \begin{array}{c}
A_{4N-2} \\
A_{4N-1} 
\end{array} \right)&=&\mathbf{M}_{4N-1}(x_{2N-1})\cdot\left( \begin{array}{c}
A_{4N} \\
A_{4N+1} 
\end{array} \right) \nonumber \quad ,
\eea
where, as long as $E<V$, the $2\times 2$ matrices of coefficients $\mathbf{M}_{2\ell}(x_\ell)$ and $\mathbf{M}_{2\ell+1}(x_\ell)$ read

\be
\mathbf{M}_{2\ell}(x_{\ell})=\left( \begin{array}{c c}
e^{ikx_{\ell}} & e^{-ikx_{\ell}} \\
ik e^{ikx_{\ell}} & -ik e^{-ikx_{\ell}} 
\end{array} \right)  \hbox{and} \quad \mathbf{M}_{2\ell+1}(x_\ell)=\left( \begin{array}{c c}
e^{-zx_{\ell}} & e^{zx_{\ell}} \\
-z e^{-zx_{\ell}} & z e^{zx_{\ell}} 
\end{array} \right) \label{randmatr1}
\ee

or

\be
\mathbf{M}_{2\ell}(x_{\ell})=\left( \begin{array}{c c}
e^{-zx_{\ell}} & e^{zx_{\ell}} \\
-z e^{-zx_{\ell}} & z e^{zx_{\ell}} 
\end{array} \right) \hbox{and} \quad \mathbf{M}_{2\ell+1}(x_\ell)=\left( \begin{array}{c c}
e^{ikx_{\ell}} & e^{-ikx_{\ell}} \\
ik e^{ikx_{\ell}} & -ik e^{-ikx_{\ell}} 
\end{array} \right) \label{randmatr2}
\ee
if $\ell$ is, respectively, even or odd.
Assuming that the amplitude $A_0$ of the incoming wave $\psi_0$ is known and that $A_{4N+1}=0$, because there is no reflection from the right boundary in the $2N$-th region, then the linear system (\ref{trmat}) corresponds to a set of $4N$ equations in the $4N$ variables $\{A_1,...,A_{4N}\}$.
Skipping the spatial dependence of the matrices to keep our notation simple, eq. (\ref{trmat}) implies

\bea
\left( \begin{array}{c}
A_{0} \\
A_{1} 
\end{array} \right)&=&\underbrace{\mathbf{M}_{0}^{-1}\cdot\mathbf{M}_{1}\cdot\mathbf{M}_{2}^{-1}\cdot\mathbf{M}_{3}\cdot...\cdot\mathbf{M}_{4N-2}^{-1}\cdot\mathbf{M}_{4N-1}}_{\mathbf{M}}\cdot \left( \begin{array}{c}
A_{4N} \\
0 
\end{array}\right)= \nonumber\\
&=&\mathbf{M}\cdot\left( \begin{array}{c}
A_{4N} \\
0 
\end{array}\right)\label{trmat2} \quad ,
\eea
where we have also defined the $2\times 2$ matrix $\mathbf{M}$.
Equation (\ref{trmat2}) leads to 

\be
A_{0}=M_{11} A_{4N} \label{trmat3}
\ee
with $M_{11}$ denoting the element on the top left corner of the matrix $\mathbf{M}$. Relation (\ref{trmat3}) allows us to rewrite eq. (\ref{J}) in the form:
\be
S=\frac{A_{4N}^*A_{4N}}{A_{0}^*A_{0}}=\frac{1}{|M_{11}|^2} .\label{J2}
\ee
The results obtained with the TM method to the model described by eq. (\ref{se2}) are illustrated in Sec. \ref{sec:sec3}.

\section{Numerical results}
\label{sec:sec3}

We characterize the presence of disorder in the sequence of barriers letting $\rho(\hat{\lambda}) d\hat{\lambda}$ be the probability distribution of the widths $\hat{\lambda}_i$ of a generic barrier, with $i =\{1,...,N\}$, to take values in a range $d\hat{\lambda}$ centered on $\hat{\lambda}$. 
In particular, we used pseudo-random generators to investigate two relevant choices for $\rho(\hat{\lambda})$. The first is the \textit{uniform} density, with support on the unit interval, while the other is the density $\rho=\sqrt{\frac{2}{\pi}}e^{-\frac{\hat{\lambda}^2}{2}}$, supported on $\mathbb{R}^+$, which is readily obtained from the gaussian density $\rho=\mathcal{N}(0,1)$ by retaining only the positive values of the $\hat{\lambda}_i$'s. Each of the two distributions is characterized by the corresponding mean $\langle \hat{\lambda}\rangle$ and variance $\hat{\sigma}^2$. \footnote{For a uniform density $\langle \hat{\lambda}\rangle =0.5$, $\hat{\sigma}^2 = 1/12$, whereas for the ``gaussian'' density defined above, with support on $\mathbb{R}^{+}$, $\langle \hat{\lambda}\rangle =2/\pi$, $\hat{\sigma}^2 = 1-2/\pi$.}. It proves useful to introduce, for both the distributions above, the \textit{realization} mean and variance, defined, 
respectively, as
\bea
\hat{\lambda}_B&=& \frac{1}{N}\sum_{i=1}^{N}\hat{\lambda}_i ,\nonumber \\
\hat{W}_N^2&=& \frac{1}{N}\sum_{i=1}^{N}(\hat{\lambda}_i-\langle \hat{\lambda}\rangle)^2 . \nonumber
\eea
In the large $N$ limit, the random variable $\hat{\lambda}_B$ converges in probability to the mean $\langle \hat{\lambda}\rangle$ (Weak Law of Large Numbers), while the random variable $\hat{W}_N^2$ converges with probability $1$ to $\hat{\sigma}^2$. 
Since we use dimensionless variables in eq. (\ref{se2}), we introduce the rescaled barrier width as:
\be
\lambda_i=c \hat{\lambda}_i, \quad  \hbox{with} \quad c=\frac{\beta}{(1+\beta) N \hat{\lambda}_B} \quad . \label{constr}
\ee 
Therefore, for any given $N$ and $\beta$, the rescaled mean
$$
\lambda_B=\frac{\beta}{(1+\beta)N}
$$ 
is no longer a random variable, and attains the same constant value independently of the chosen probability density. On the other hand, the rescaled realization variance 
$$
W_N^2=\frac{c^2}{N}\sum_{i=1}^{N}(\hat{\lambda}_i-\hat{\lambda}_B)^2
$$
remains a random variable which, for large $N$ converges a. e. to $\sigma^2=c^2 \hat{\sigma}^2$. We introduce the vector-valued random variable $\Lambda_N$, defined by 
\be
\Lambda_{N}=\left\{\lambda_1,...,\lambda_N \right\} \quad , \label{Lambda}
\ee 
which corresponds to a given realization of the sequence of barriers and will be referred to as a \textit{microscopic configuration}.

%\begin{figure}
%   \begin{center}
%   \includegraphics[width=7cm]{microcan.eps}
%  \hspace{2mm} \
%\includegraphics[width=7cm]{gauss.eps}
%   \caption{\textit{Left panel}: Uniform probability density, with values $\tilde{\lambda}_i \in[0,1]$. \textit{Right panel}: Probability density derived from the gaussian density $\mathcal{N}(0,1)$, supported on the positive branch of the real line.}\label{density}
%   \end{center}
%\end{figure}

For given $\beta$ and $N$, one may, then, consider the collection $\Omega=\{\Lambda_{N}^{(1)},...,\Lambda_{N}^{(N_r)}\}$ of $N_r$ random realizations of the sequence of barriers which have been constructed numerically. 
Then, the average of a random observable $\mathcal{O}$ over the sample $\Omega$  writes: 
\be
\langle \mathcal{O} \rangle_{\Omega}=\frac{1}{N_r}\sum_{\mu=1}^{N_r}\mathcal{O}\left(\Lambda_{N}^{(\mu)}\right) \label{ranav}
\ee
Among the possible configurations, the regular one
\be
\Lambda_{N}^{B}=\left\{\lambda^B,...,\lambda^B \right\} \label{Bloch} \quad,
\ee
which approximates the \textit{infinite superlattice} of the literature on Bloch waves \cite{mermin,harris}, will be crucial also in our work. 
In our numerical simulations we set $\beta=0.1$ and we investigated the behavior of the coefficient $S(N,\Lambda_N,V,E;T)$  at a given temperature $T$, as a function of the number of barriers $N$, of the microscopic configuration $\Lambda_N$, of the (dimensionless) energy of the barriers $V$ and of the (dimensionless) energy $E$. 

\begin{figure}
\centering
\includegraphics[width=9cm]{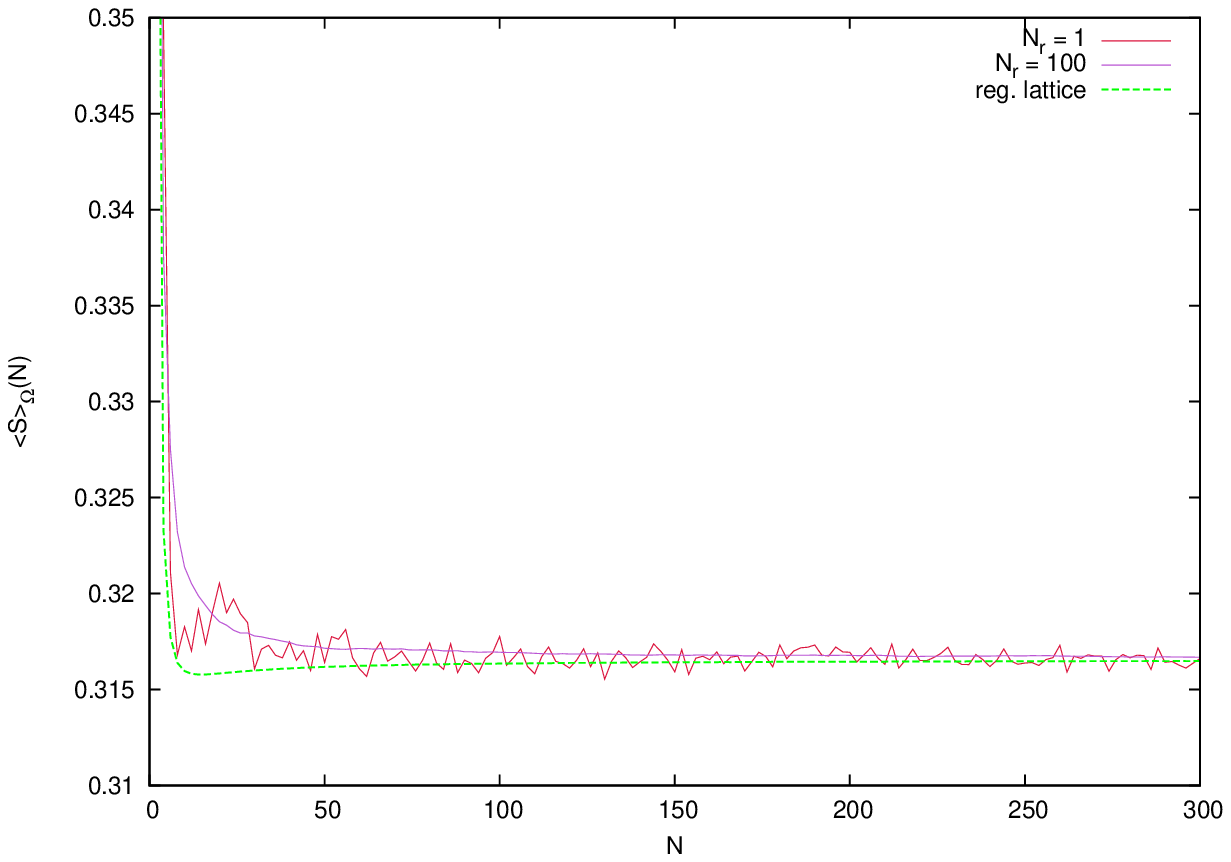}
\vspace{2mm}
\includegraphics[width=9cm]{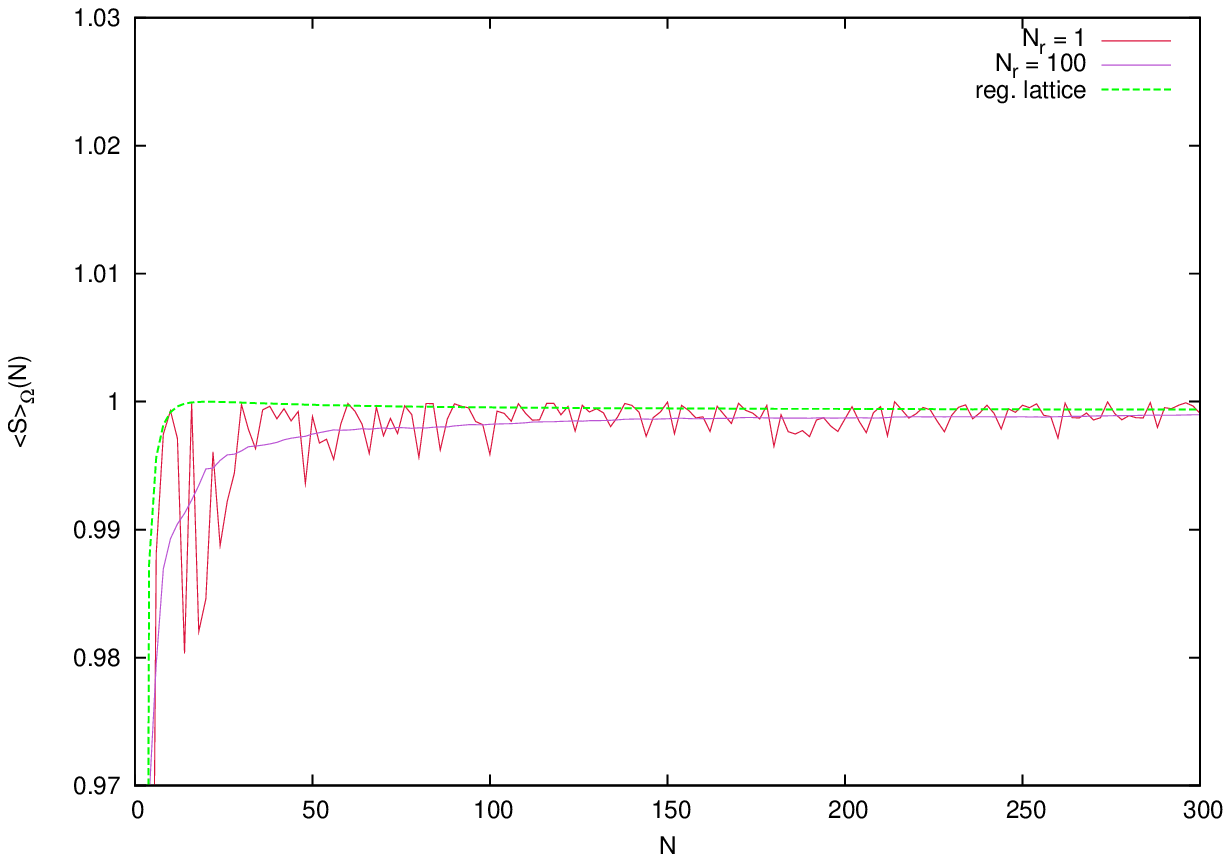}
\caption{\textit{Top panel}: Behavior of $S(N,\Lambda_N,V,E;T)$ vs. $N$, for $V=20$, $E=2.0$ at $T=300 K$ and for different microscopic configurations: the red curve corresponds to a single uniformly distributed configuration $\Lambda_N$, the purple curve represents the average $\langle S\rangle_\Omega$ over an ensemble of uniformly distributed configurations, while the dashed green curve corresponds to the realization (\ref{Bloch}).
\textit{Bottom panel}: Same curves as those shown in the top panel, but here evaluated at $E=13.8$, where a condition of tunneling resonance is realized.}
\label{Eqmesk}
\end{figure}

\begin{figure}
   \centering
\includegraphics[width=9.5cm]{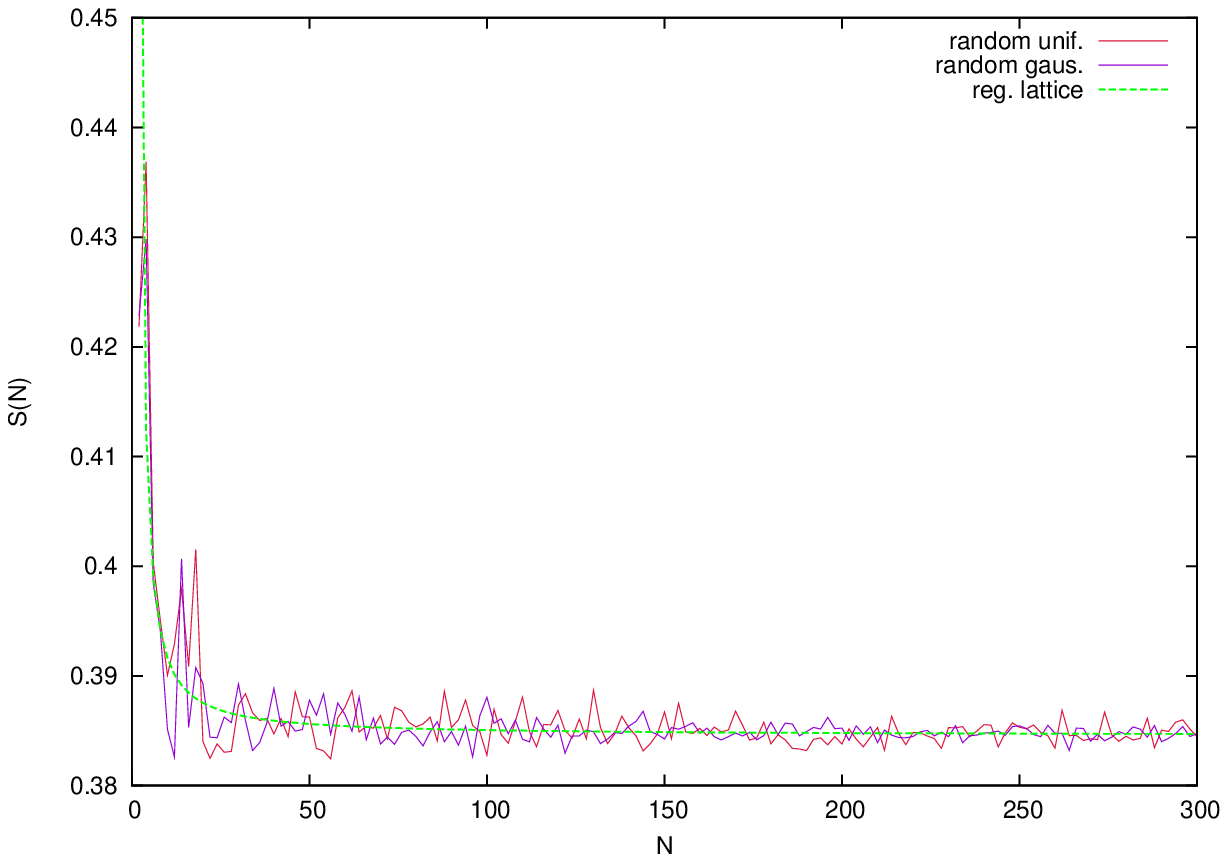}
  \vspace{2mm} 
   \includegraphics[width=9.5cm]{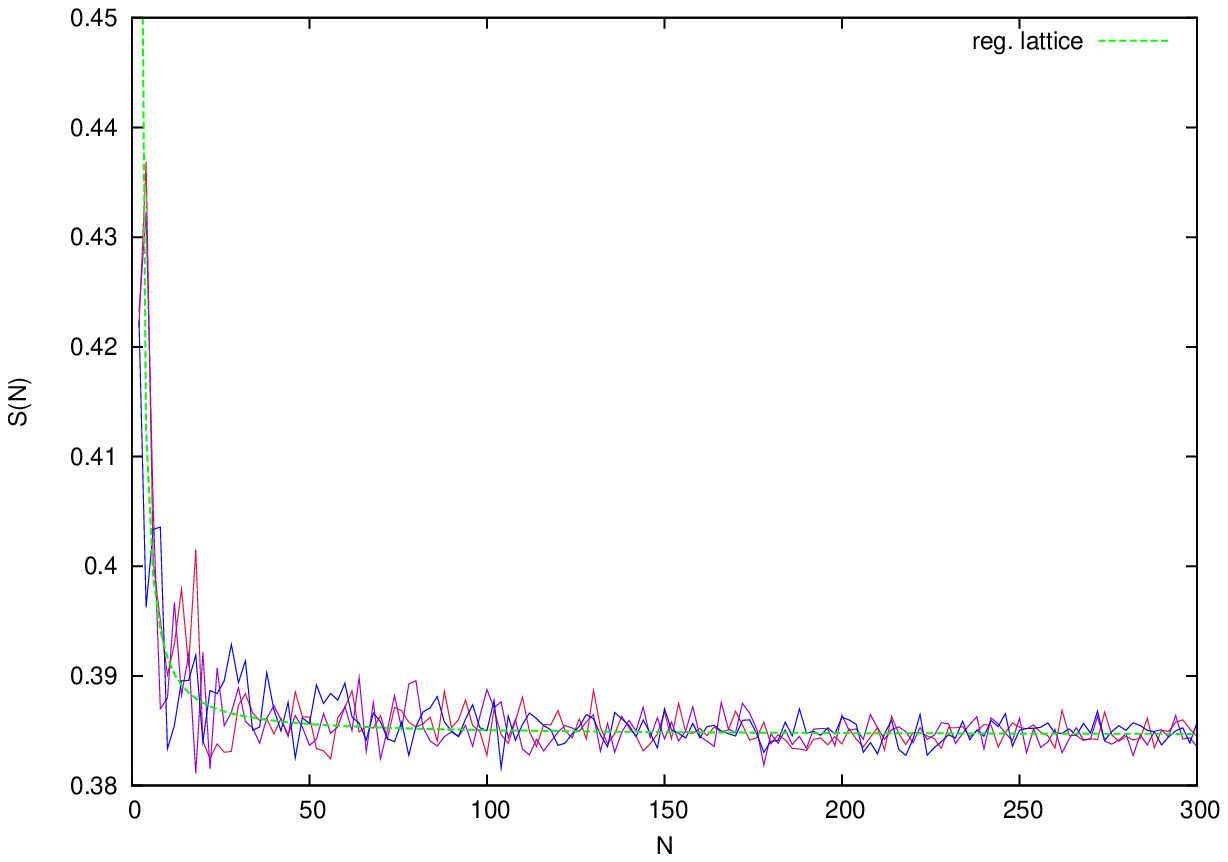}
   \caption{\textit{Top panel}: Behavior of $S(N,\Lambda_N,V;T)$ for $V=20$ and at $T=300 K$ corresponding to three different microscopic configurations: the regular lattice  configuration $\Lambda_N^B$ (thick green dashed curve), a uniformly distributed configuration (red curve) and a normally distributed one (blue curve). \textit{Bottom panel}:Behavior of $S(N,\Lambda_N,V;T)$ vs. $N$, for $V=20$, $T=300 K$ and for three different random uniformly distributed configurations $\Lambda_{N}^{(\mu)}$, with $\mu=1,2,3$. All curves tend to overlap with that corresponding to the regular lattice.}\label{noise1}
\end{figure}

Figure \ref{Eqmesk} illustrates the behavior of $S(N,\Lambda_N,V,E;T)$, $S(N,\Lambda_{N}^{B},V,E;T)$ and of the average $\langle S\rangle_\Omega$ over an ensemble characterized by $\rho(\hat{\lambda})=1$ at $T=300 K$ and for two different values of the wave energy $E$, one of which leads to a condition of resonant tunneling \cite{raz}.
The plots reveal that the values of $S$ and of its average $\langle S\rangle_\Omega$ tend, for growing $N$, to the same value, as intuitively expected. However, it is interesting to note that this value coincides with the most probable value of $S(N,\Lambda_N,V,E;T)$, which is the value corresponding to the regular lattice configurations (\ref{Bloch}), hereafter denoted as $S_B(N,V,E;T)$. This holds for all values of $V$ and of $E$. Let us now consider a thermostat located at the left boundary, so that the plane waves entering the bulk have an energy obeying a classical equilibrium distribution at a given temperature $T$.
In the following plots we consider a one dimensional Maxwellian probability density
\be
f_{eq}(E)=\sqrt{1/(\pi E)} e^{-E} \nonumber
\ee 
and we average over all energies to obtain 
\be
 S(N,\Lambda_N,V;T)=\int_{0}^{\infty} S(N,\Lambda_N,V,E;T) f_{eq}(E) dE \label{eqaver} 
\ee
where the coefficient $S$, integrated in the r.h.s. of eq. (\ref{eqaver}), is computed according to eq. (\ref{J2}). 
The typical behavior of $S(N,\Lambda_N,V;T)$ as a function of the number of barriers $N$, for $V=20$ is illustrated in Fig. \ref{noise1}, which refers to three different microscopic configurations: the regular lattice $\Lambda_N^B$  introduced in eq. (\ref{Bloch}) (red curve), a random uniform configuration (green curve), and a ``gaussian'' distribution (blue curve).

\begin{figure}
   \centering
\includegraphics[width=9.5cm]{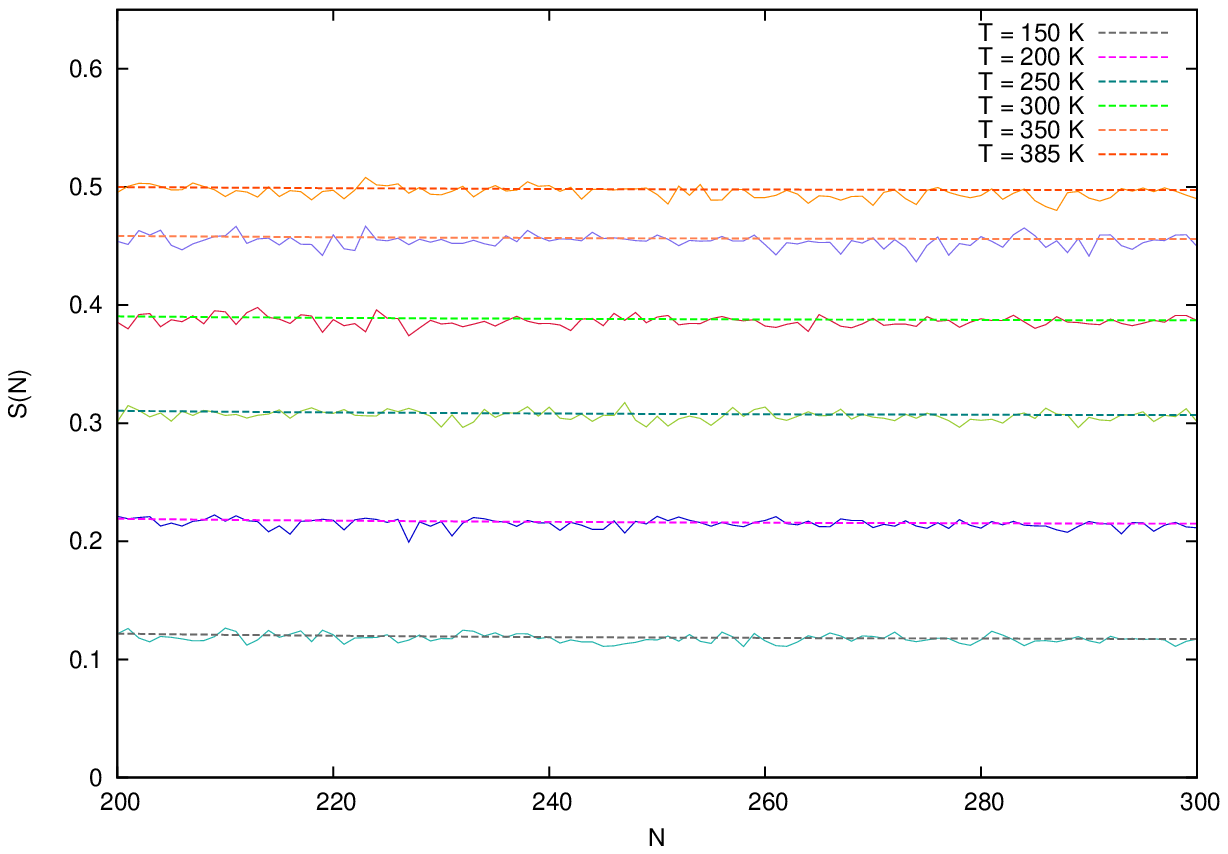}
  \vspace{2mm} 
   \includegraphics[width=9.5cm]{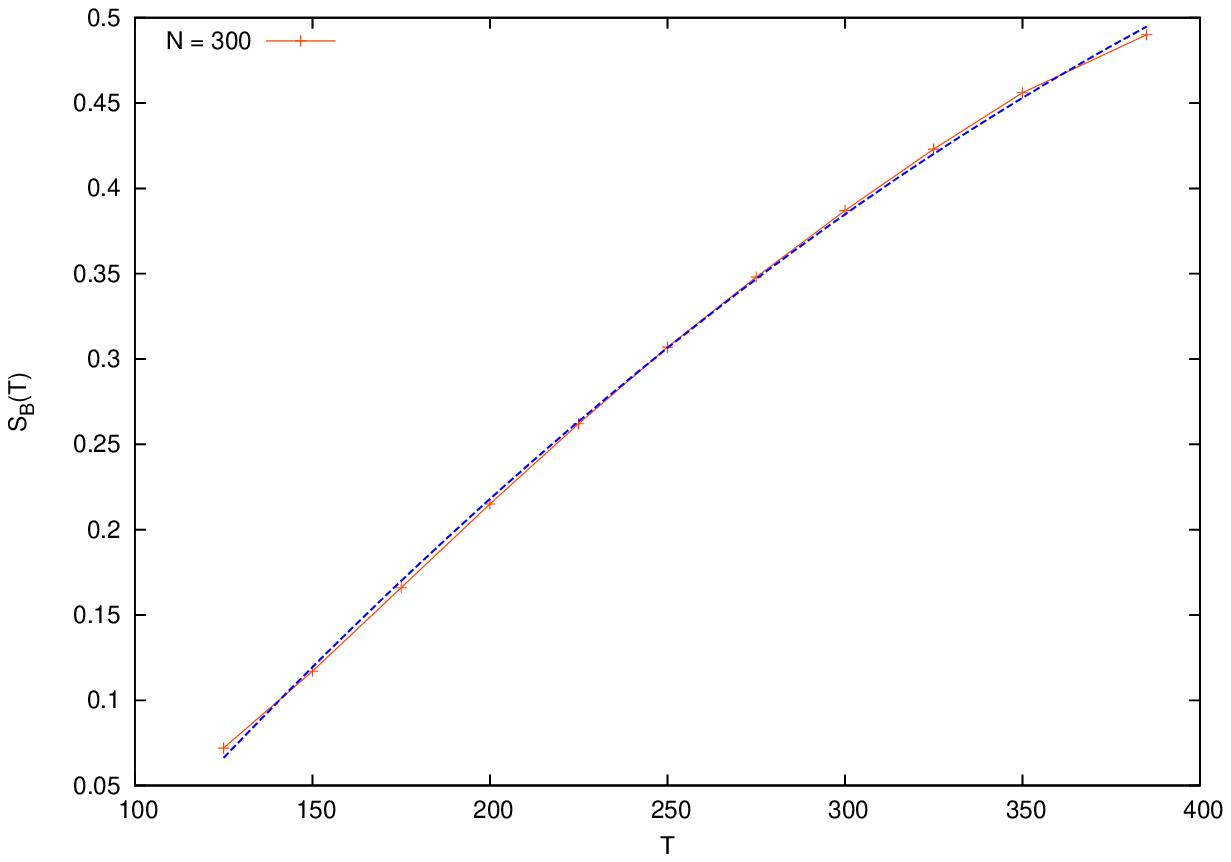}
   \caption{\textit{Top panel}: Values for the observable $S(N,\Lambda_N,V;T)$ at different temperatures, with $V=20$. Shown are the numerical results for uniformly distributed configurations (straight lines) and for the corresponding regular lattice  configurations $\Lambda_N^B$ (thick dashed curves). \textit{Bottom panel}: Fitting of the numerical data of $S_B(N,\Lambda_N,V)$, calculated at $N=300$ and with $V=20$, at different temperatures with the curve $\phi(T)=aT^2+bT+c$, with $a=-2.022\pm 1.891\cdot10^{-7}$, $b= 0.002\pm 9.672\cdot10^{-5} $ and $c=-0.237\pm 0.011$.}\label{temperature}
\end{figure}

The result illustrated in the top panel of Fig. \ref{noise1} shows that, for large $N$, the values of $S(N,\Lambda_N,V;T=300)$ corresponding to single random realizations fluctuate around $S_B(N,V;T=300)$ for both distributions.
Similarly, the lower panel of Fig. \ref{noise1} shows the behavior of $S(N,\Lambda_N,V;T)$ vs. $N$ for three different random uniform configurations.\\
We also investigated the dependence of $S(N,\Lambda_N,V;T)$ on $T$. The top panel of Fig. \ref{temperature} corroborates, in the limit of large $N$, the numerical results illustrated in the top panel of Fig. \ref{noise1}. Namely, the trend of the random values $S(N,\Lambda_N,V;T)$ to approach $S_B$ persists even at lower and higher temperatures. This is one of the prominent features of our model. The lower panel of Fig. \ref{temperature} shows that the variation of $S_B$, over a wide range of temperatures, can be suitably fitted by a quadratic curve, $\phi(T)=aT^2+bT+c$.

Figure \ref{noise1} anticipates two further crucial aspects which will be addressed in more detail below. The first concerns the magnitude of the fluctuations of the values of $S(N,\Lambda_N,V;T)$, $\sigma_\rho(N,V)=\sqrt{\langle (S-\langle S\rangle_\Omega)^2\rangle_\Omega}$, which decreases with $N$. This decay of the size of the fluctuations allows us to identify a ``mesoscopic'' scale $N_{meso}$. The number $N_{meso}$ depends on $V$ and the coefficient $S(N,\Lambda_N,V;T)$ depends only weakly on the microscopic configuration, if $N\ge N_{meso}$.
The second scale $N_{macro}>N_{meso}$ is such that $S(N,\Lambda_N,V;T)$ depends neither on the configuration nor on the number of barriers if $N\ge N_{macro}$. In other words, $S$ attains a value which is, practically, that concerning infinitely finely structured samples at a finite fixed length. Note that $N_{macro}$ depends also on $V$.
 
\begin{figure}
   \centering
   \includegraphics[width=0.8\textwidth]{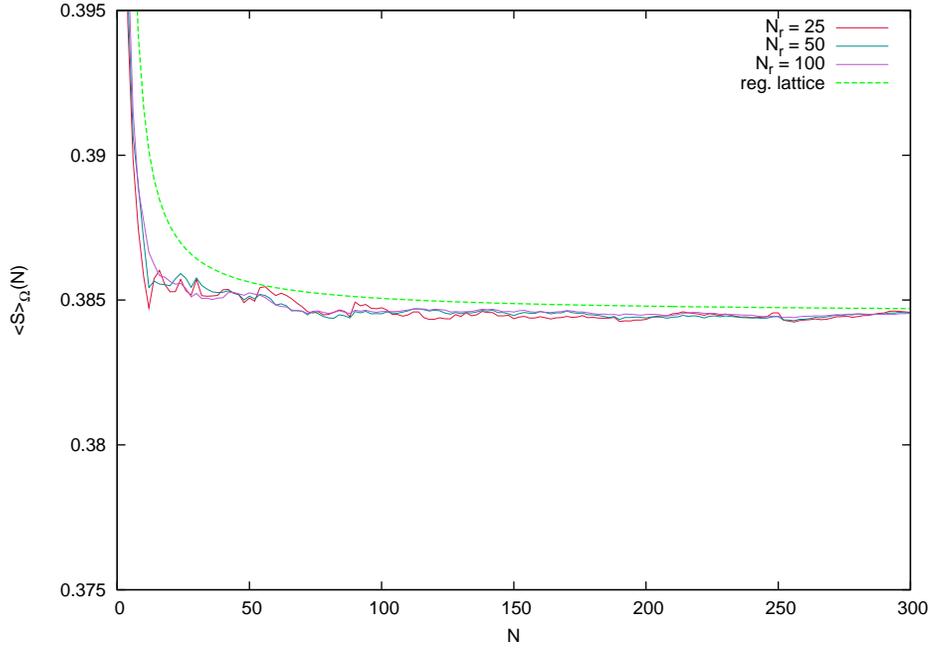}
   \caption{Behavior of $\langle S\rangle_{\Omega}$  vs. $N$, for $V=20$, $T=300 K$ and for different values of $N_r$.}\label{noise2}
\end{figure}
Next, in Fig. \ref{noise2}, we plotted the behavior of $\langle S\rangle_\Omega(N,V;T)$ with $N$, for $V=20$, and we compare it with the value of $S_B(N,V;T)$ for regular lattices. The result is consistent with those of Figs. \ref{Eqmesk} and \ref{noise1}, in that it shows the regime corresponding to $N\gg 1$, where the curve of $\langle S\rangle_\Omega(N,V;T)$ varies very slowly with $N$ and approaches the value $S_B(N,V;T)$. This convergence process is a collective effect, in that it does not depend on the underlying disorder and is characterized by the delocalization of the wave function typical described by the Bloch waves theory \cite{mermin} for regular lattices. The plot in Fig. \ref{noise2} reveals that, given an ensemble $\Omega$ of random uniform configurations, one has:
\be
\langle S\rangle_\Omega(N,V;T)\simeq S_B(N,V;T)\quad  \hbox{for large $N$}\label{equiv} \quad .
\ee 
Indeed Fig. \ref{noise2} shows that $N\geq200$ suffices to reach a good accuracy even with small samples $\Omega$.\\
Moreover, we verified that the energies of the incoming wave, at a given temperature, lie in the conducting band of the infinite periodic chain of barriers, which implies that $S_B(N,V;T)>0$.
A few comments are in order, here.
In the first place, the absence of localization can be traced back to the fixed finite amount of insulating material, which we have even in the $N\rightarrow\infty$ limit, because $L$ is fixed. As a consequence, incoming waves may, at most, be damped by a finite factor, except, perhaps, for a negligible set of energies which we have not observed. This distinguishes our model from the tight-binding model, which is more extensively investigated in the specialized literature, and also prevents the application of the Furstenberg's theorem \cite{Vulp}. Indeed, introducing the $(N+1)$-th barrier in one our systems amounts to a complete rearrangement of the previous $N$ barriers. Mathematically, this means that the product of the first $N$ random matrices is replaced by a new product. Differently, the case of ergodic-like theorems, such as Furstenberg's one, applies to products of $N$ random matrices which are multiplied by the $(N+1)$-th matrix. Moreover, the decrease of the size of fluctuations with $N$, which will be explored in more detailed below, can be regarded as a phenomenon of self-averaging of the observable $S$, which is quite common \cite{Vulp}.
However, our results, further supported by the analysis of the PDF of the transmission coefficient, Fig. \ref{prob},\ref{ratefunct0} below, show that the average over the disorder of the random values $S(N,\Lambda_N,V;T)$ attains the specific value $S_B(N,V;T)$ pertaining to the regular configuration.
In other words, the transmission coefficient self-averages around a value $\langle S \rangle_\Omega(N,V;T)$ which is close to $S_B$ even for small $N$ and tends to $S_B$ in the $N\rightarrow \infty$ limit. As shown in Fig. \ref{ratefunct0}, given a sample $\Omega$ of uniform realizations, $S_B(N,V;T)$ corresponds to the most probable value of the random variable $S$ in the sample, which, when $N$ grows, tends also to the mean $\langle S\rangle_\Omega(N,V;T)$.
Furthermore, being obtained from the dimensionless SE (\ref{se2}), this result is independent of the length of the system (as long as $L$ is fixed) and does not result in localization effects when $L$ is large, because $S_B$ does not vanish.\\
Let us now investigate, more accurately, the structure of the fluctuations, in the sample $\Omega$ of random uniform configurations at temperature $T=300 K$.
Denote by $\rho_N(S)dS$ the probability that $S$ falls in the interval $dS$, so that:
\be
\langle S \rangle_\Omega=\int S \rho_N(S) dS  \quad .\label{rho}
\ee
The numerical results presented so far on the relation between $\langle S \rangle_\Omega(N,V;T)$ and $S_B(N,V;T)$ and their mutual relation, eq. (\ref{equiv}), as well as on the decrease of the fluctuations size with growing $N$, indicate that $\rho_N(S)$  peaks more and more around the reference value $S_B(N,V;T)$. 
To show this, we numerically calculated the quantity $\rho_N(S)$ from eq. (\ref{ranav}) and we plotted in Fig. \ref{prob} the resulting curves for different values of $N$ and for $V=20$. The maxima of the PDF in Fig.\ref{prob} are located at $S=S_B(N,V;T)$, cf. also Fig. \ref{ratefunct0}, and tend, for large $N$, to the mean values $\langle S \rangle_\Omega (N,V;T)$.
One further realizes that $\rho_N(S)$ obeys a large deviation principle with a given rate functional $\zeta_N$, in the sense that the limit
\be
\lim_{N\rightarrow \infty}\frac{-\log{\rho_N(S)}}{N}=\lim_{N\rightarrow \infty} \zeta_N(S)=\zeta(S) \label{largedev}
\ee
exists.
Figure \ref{ratefunct0} shows, for the range in which we have good statistics, that the large deviation rate functional $\zeta$ is apparently smooth and strictly convex. It is worth pointing out, again, that the $N\rightarrow \infty$ limit is not achieved in the standard fashion of products of random matrices.

\begin{figure}
\begin{center}
  \includegraphics[width=12.5cm]{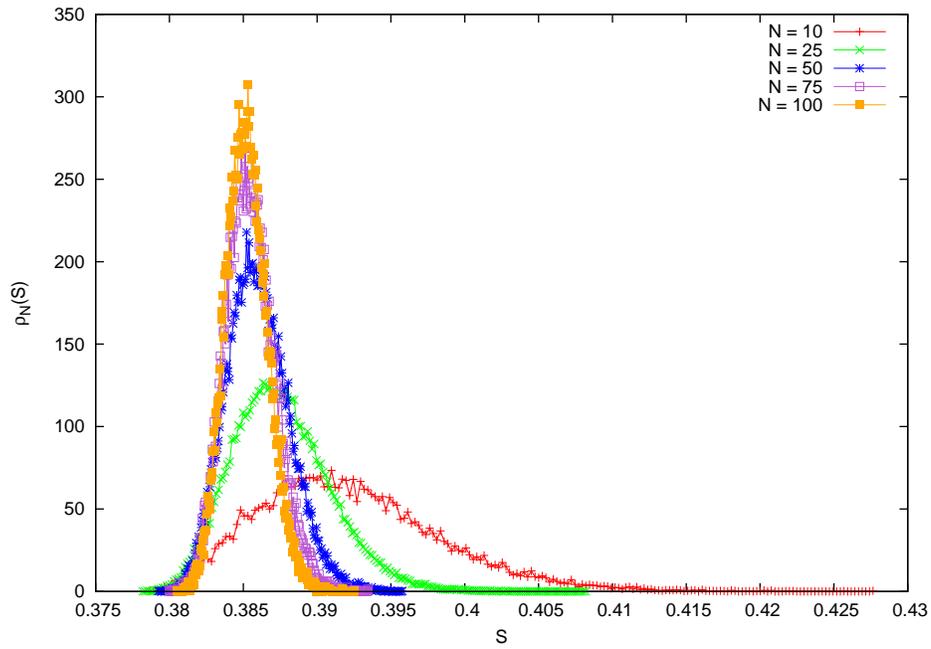}\\
  \caption{Probability densities $\rho_N(S)$ for different values of $N$ and for $V=20$.}\label{prob}
   \end{center}
\end{figure}

\begin{figure}
\begin{center}
  \includegraphics[width=0.8\textwidth]{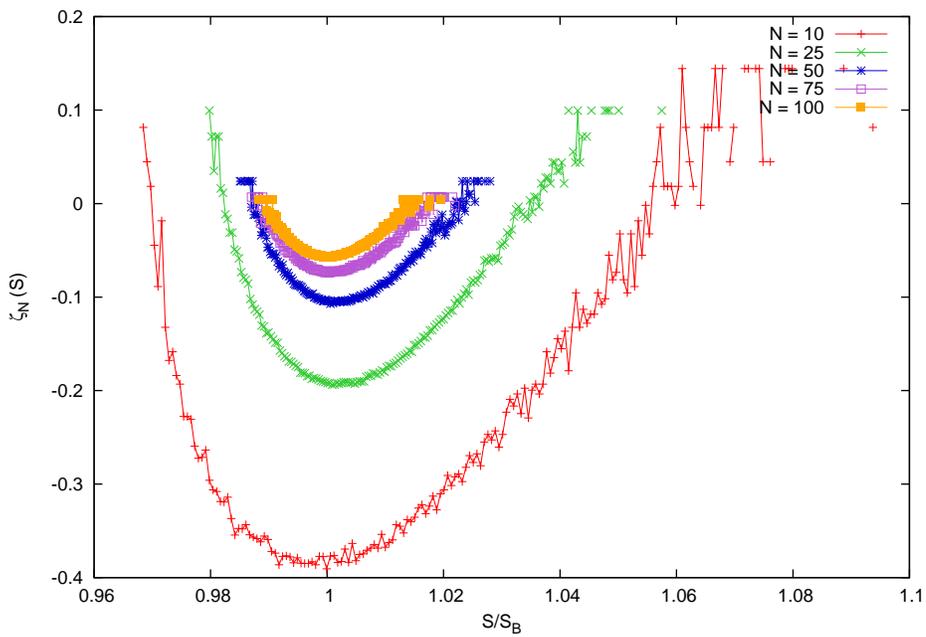}\\
  \caption{Rate functional $\zeta_N$ associated to $\rho_N(S)$ for different values of $N$ and for $V=20$. The curves $\zeta_N$ move downwards for growing $N$, so that, as expected, in the $N\rightarrow \infty$ limit, $\zeta(S)$ intersects the horizontal axis only in $S=S_B$.}
\label{ratefunct0}
   \end{center}
\end{figure}

\begin{figure}
\begin{center}
  \includegraphics[width=0.8\textwidth]{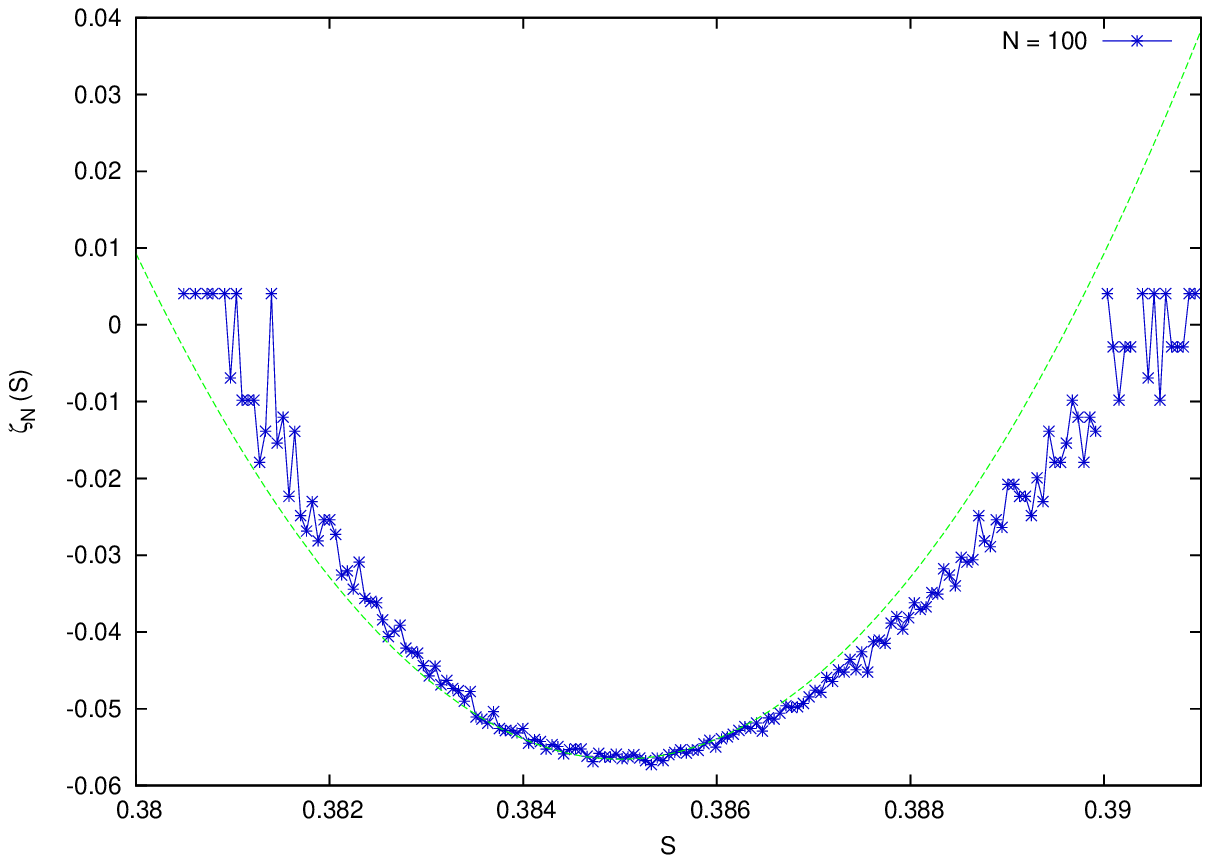}\\
  \caption{Rate functional $\zeta_N(S)$ associated to $\rho_N(S)$. \textit{Blue points}: results of the numerical simulation with $N=100$ and $V=20$. \textit{Green dashed line}: Fitting of numerical data with the parabola $a(S-S_B(N))^2+b$, with parameters $a=2635.5 \pm 82.96$ and $b=-5.655\cdot10^{-2}\pm 3.38\cdot10^{-4}$.}\label{EQclt}
   \end{center}
\end{figure}

\begin{figure}
\begin{center}
  \includegraphics[width=0.8\textwidth]{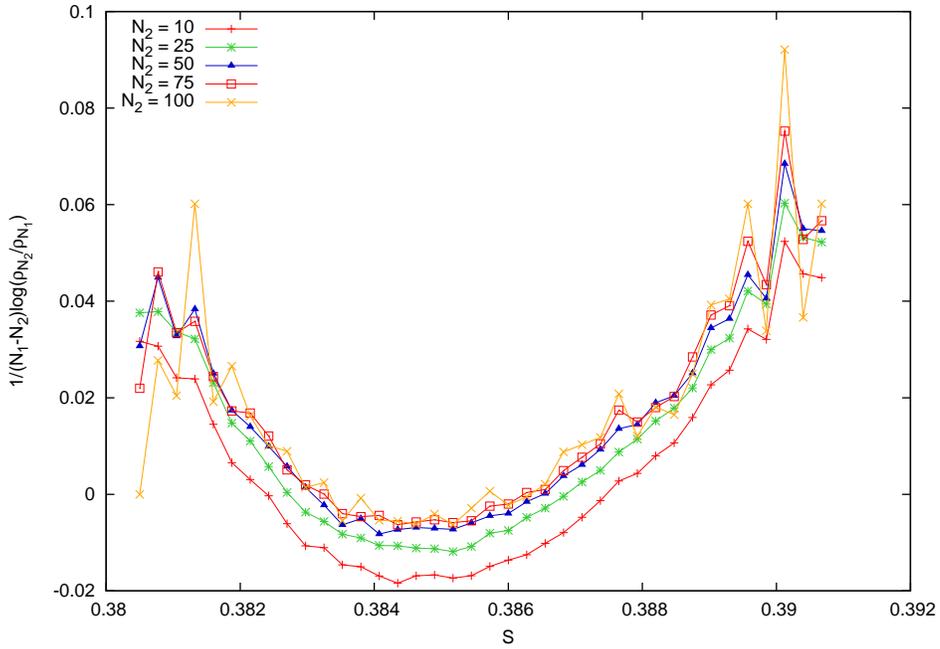}\\
  \caption{Plot of the function $\frac{1}{N_1-N_2}\log\left(\frac{\rho_{N_2}}{\rho_{N_1}}\right)$ with $N_1=125$ and $N_2 = \{10,25,50,75,100\}$.}\label{ratefunct1}
   \end{center}
\end{figure}

It is also interesting to note, in Fig. \ref{ratefunct0}, that the locus of the minima, i.e. of highest probability density, of all the curves $\zeta_N(S)$ is represented by $S_B(N,V;T)$.
Moreover, Fig. \ref{EQclt} reveals that, for small deviations from $S_B(N,V;T)$, the rate functional $\zeta_N(S)$ is quadratic, as expected where the central limit theorem applies. 
Following the approach outlined in Ref. \cite{RonMor}, if a large deviation functional $\zeta(S)$ exists, for any finite $N$ one may write:
\be
\zeta(S)=\frac{1}{N}\log\rho_N(S)+z_N(S) \label{rond1}
\ee
where $z_N$ is a correction term expected to be of order $O(\frac{1}{N})$ as long as correlations decay sufficiently fast, cf. \cite{artuso, melb}. This correction term may be further split as $z_N(S)=\phi(S)/N+\hat{z}_N(S)$.
Thus, if we consider two different values of $N$, say $N_1$ and $N_2$, one may write \cite{RonMor}:

\be
\zeta(S)=\frac{1}{N_1-N_2}\left[\log(\rho_{N_2}/\rho_{N_1})+N_2 \hat{z}_{N_2}-N_1 \hat{z}_{N_1}\right] \label{rond2}
\ee
where the remaining correction terms $\hat{z}_{N_i}$ are expected to give a smaller contribution to $\zeta$ than $z_{N}$ in eq. (\ref{rond1}) does.
We numerically verified the validity of the finite $N$ estimate of the large deviation functional $\zeta(S)$, given by eq. (\ref{rond2}), observing the rapid collapse of all curves of Fig. \ref{ratefunct0} into a single curve, in Fig. \ref{ratefunct1}.

\begin{figure}
   \centering
\includegraphics[width=8.5cm]{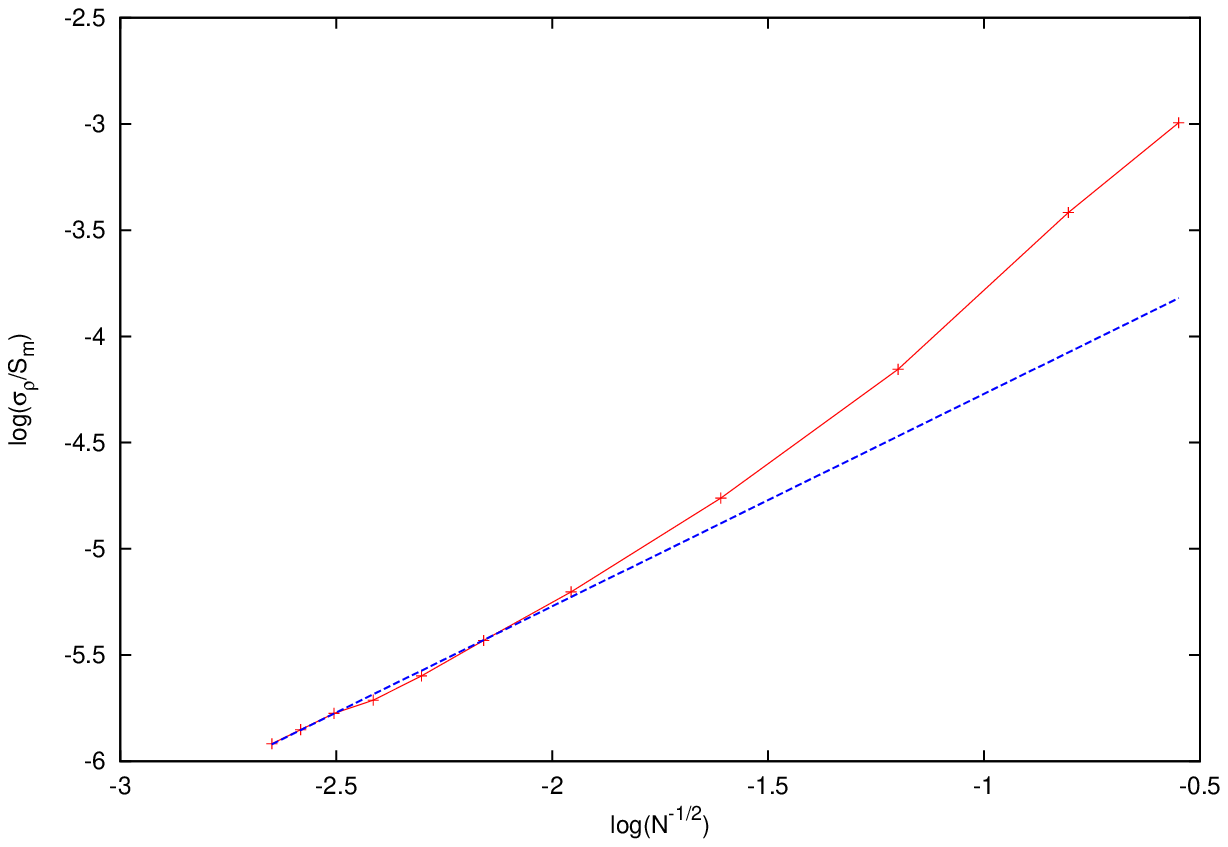}
\vspace{2mm}
\includegraphics[width=8.5cm]{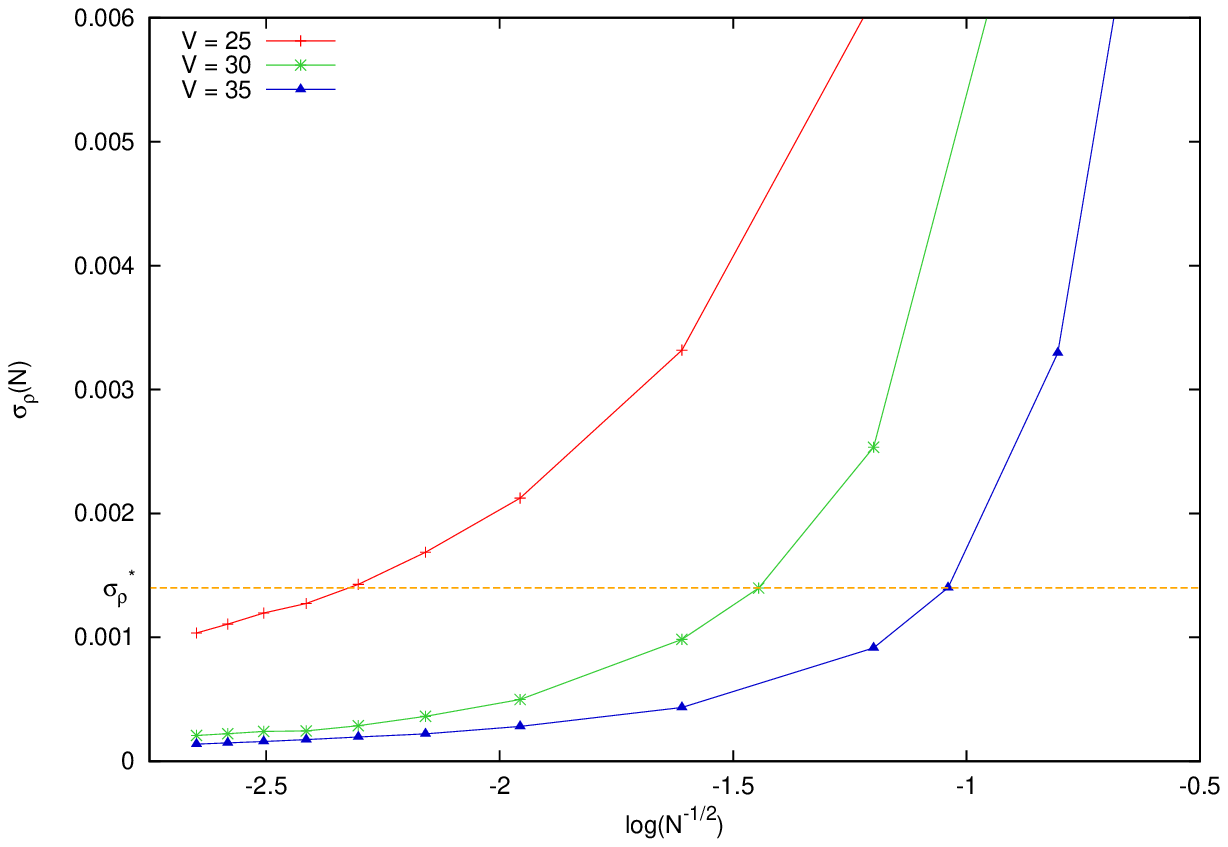}
   \caption{\textit{Top panel}: Fit of $\sigma_\rho/S_m$ (red curve), evaluated for $V=20$, with the curve $f(N)=a\times N^{-\frac{1}{2}}$, with $a=3.798383\times 10^{-2} \pm 4.707 \times 10^{-4}$. The plot confirms the asymptotic behavior of $\sigma_\rho/S_m$, given by eq. (\ref{thermlim2}). \textit{Bottom panel}: Behavior of $\sigma_\rho(N,V)$  vs. $N$, for $V=\{25,30,35\}$. The thick orange dashed line denotes the value $\sigma_\rho^*$ below which we may conventially assume that the size of fluctuations is practically negligible.}\label{EQvariance}
\end{figure}

From the validity of a principle of large deviations and of the central limit theorem, one expects the following asymptotic behavior for the fluctuations of the transmission coefficient:

\be
\frac{\sigma_\rho(N,V)}{S_m(N,V)}\sim \left(\sqrt{N}\right)^{-1} \quad , \quad  \hbox{for $N\gg 1$}. \label{thermlim2}
\ee
We numerically checked that this is the case by evaluating the ratio in eq. (\ref{thermlim2}), cf. the top panel of Fig. \ref{EQvariance}, for $V=20$. The bottom panel of Fig. \ref{EQvariance} further illustrates the decay of  $N_{meso}$ with $V$: taking $N_{meso}$ such that $\sigma_\rho(N,V)<\sigma_{\rho}^*=1.4\times 10^{-3}$ $\forall N>N_{meso}$, where $\sigma_{\rho}^*$ is considered small, $N_{meso}(V)$ is found to rapidly decrease with $V$: we obtain $N_{meso}=100,18,8$ for, respectively, $V=25,30,35$. Then, $N_{meso}$ tends to $0$ when the barrier height diverges: in this limit the transmission coefficient vanishes for any configuration $\Lambda_N$, hence the fluctuations are suppressed. 

\section{Concluding remarks}

In this work we investigated a quantum mechanical one-dimensional system, in equilibrium with a classical heat bath. The role played by the 
disorder on the transmission coefficient $S$ has been investigated.
Our numerical investigation sheds light on the existence of appropriate scales, respectively delimited by 
$N_{meso}$ and $N_{macro}$, which are not as widely separated as in thermodynamic systems.
The main result of our work suggests that, in presence of an external heat bath and off-diagonal disorder, 
the wave function is delocalized over all the system length. This absence of localization reflects the long range correlations which extend over the whole length of our systems and prevent the low 
tunneling limit. Therefore, the effective tri-diagonal Hamiltonian for given bound states in a potential 
well, which is common to studies of the Anderson model, does not
pertain to our systems and should in case be replaced by a full matrix. The presence of correlations is
known to lead to extended states in models enjoying off-diagonal disorder, cf Ref. \cite{Izrailev,zhang}. Furthermore, our procedure for the $N\rightarrow\infty$ limit makes the Furstenberg theorem inapplicable. The 
variable $S$ is self-averaging for growing $N$, and tends to the most probable value. Interestingly, 
this value is $S_B$, which corresponds to a microscopic ordered array of barriers and wells.
This observation has technological implications, as it suggests the possibility of predicting tunneling 
resonances for off-diagonal disordered systems in contact with an external bath, by investigating the 
corresponding ordered system, which is amenable to an analytical treatment. Moreover, we have numerically 
observed the validity of a principle of large deviations for the probability density $\rho_N(S)$, extending the results presented in Ref. \cite{SE}, where, finally, we verified that the most probable 
value of the transmission coefficient corresponds to the regular configurations.

\section{Acknowledgements}

M.C. wishes to thank Giuseppe Luca Celardo, Alberto Rosso, Alain Comtet and Christophe Texier for useful discussions. 
L.R. gratefully acknowledges financial support from the European Research Council under the European
Community's Seventh Framework Programme (FP7/2007-2013) / ERC grant
agreement n 202680.  The EC is not liable for any use that can be made
on the information contained herein.

\end{document}